# Time-varying effect in the competing risks based on restricted mean time lost

Zhiyin Yu[1], Zhaojin Li[1], Chengfeng Zhang[1], Yawen Hou[2],

Derun Zhou[1], Zheng Chen[1*]

1. Department of Biostatistics, School of Public Health, Southern Medical University, Guangzhou, China

2. Department of Statistics and Data Science, School of Economics, Jinan University, Guangzhou, China

* Correspondence to: Zheng Chen, Email: zheng-chen@hotmail.com.

**Abstract**: Patients with breast cancer tend to die from other diseases, so for studies that focus on breast cancer, a competing risks model is more appropriate. Considering subdistribution hazard ratio, which is used often, limited to model assumptions and clinical interpretation, we aimed to quantify the effects of prognostic factors by an absolute indicator, the difference in restricted mean time lost (RMTL), which is more intuitive. Additionally, prognostic factors may have dynamic effects (time-varying effects) in long-term follow-up. However, existing competing risks regression models only provide a static view of covariate effects, leading to a distorted assessment of the prognostic factor. To address this issue, we proposed a dynamic effect RMTL regression that can explore the between-group cumulative difference in mean life lost over a period of time and obtain the real-time effect by the speed of accumulation, as well as personalized predictions on a time scale. Through Monte Carlo simulation, we validated the dynamic effects estimated by the proposed regression having low bias and a coverage rate of around 95%. Applying this model to an elderly early-stage breast cancer cohort, we found that most factors had different patterns of dynamic effects, revealing meaningful physiological mechanisms underlying diseases. Moreover, from the perspective of prediction, the mean C-index in external validation reached 0.78. Dynamic effect RMTL regression can analyze both dynamic cumulative effects and real-time effects of covariates, providing a more comprehensive prognosis and better prediction when competing risks exist.

personalized prediction

# 1 Background

Elderly patients with early-stage breast cancer (particularly when comorbidities are advanced) tend to die from other diseases. That is, the number of deaths from non-breast cancer is large[1]. Patients may experience a variety of outcomes: death from breast cancer, death from heart disease and so on. Assuming the outcome we are interested in is death from breast cancer (event of interest), then we hope to observe the time from the start of follow-up to the occurrence of the event of interest, but this will not be observed for the patient dying from heart disease (competing event). In this case, traditional single-endpoint survival analysis, such as the Cox proportional hazards model, only considers the event of interest and treats patients who die of heart disease as censored simply. However, this doesn't meet the non-informative censoring hypothesis. That is, the risk of dying from breast cancer among women who have already died of heart disease needs to be the same as that among women who remain in follow-up. However, patients who have already died of heart disease will not die of breast cancer again. So it is improper to simply treat patients who experience competing events as censored, which will overestimate the cumulative incidence of the event of interest and result in bias[2–4]. To solve the situation where multiple outcomes compete with each other, we should consider competing risks models.

In traditional multivariate analysis of competing risks, cause-specific Cox regression and Fine-Gray regression are often used, and the corresponding effect sizes are the cause-specific hazard ratio (cHR) and subdistribution hazard ratio (sHR). However, both the cHR and sHR are relative indicators, defined as the ratio of the hazard function. It is difficult for clinicians to interpret them as intuitive clinical benefits and communicate with patients[5,6]. For example, when the sHR of estrogen receptor (ER) is 0.43, that is, the risk of death in the ER-positive group is 0.43 times that in the ER-negative group, and because the baseline hazard is unknown generally, the absolute risk of death of two groups cannot be known. In addition, both cause-specific Cox regression and Fine-Gray regression models need to satisfy the proportional hazards assumption. From a clinical perspective, clinicians or patients are more interested in direct (absolute) effect sizes on a time scale, e.g., how long will I live? How long will surgery extend my life expectancy? As Blagoev[6] points out, "While a hazard ratio has some value, for the clinician caring for a patient and, more importantly, the patient, it does not convey benefit in terms that are meaningful—how much longer will the patient live or live without experiencing disease progression." Therefore, restricted mean time lost (RMTL) has been proposed as an alternative measure to the hazard ratio[7–12]. The RMTL is the area under the cause-specific cumulative incidence function (CIF) over a period of time (from 0 to a restricted time point $\tau$), which can be interpreted as the life expectancy lost due to a specific cause in this period of time. Compared with the hazard ratio, the interpretation of RMTL is more intuitive, giving the life lost of each group due to death from breast cancer over a period of time and measuring the effect of a factor by the difference in RMTL between groups. For example, Figure 1 shows areas under the CIF for patients in the ER-positive group and ER-negative group. During the 10.5 years ($\tau = 10.5$), the mean life lost due to death from breast cancer was 0.9 ($S_0$) years for patients in the ER-positive group and 2.4 ($S_0 + S_1$) years for patients in the ER-negative group, and ER-negative patients lost an additional 1.5 (2.4-0.9) years of life on average. At the same time, RMTL does not need to meet the proportional hazards assumption.

The existing multivariate analysis of RMTL includes two methods: regression based on the pseudo-value method[11] and regression based on inverse probability of censoring weighting(IPCW)[12]. The regressions mentioned above only concern the cumulative effects of prognostic factors during a $\tau$-year follow-up, which are constant values (static or time-fixed effects). However, the real-time effects of many covariates (eg: covariates with time-varying effects) vary, for example, the real-time effect of chemotherapy tends to decrease with increasing follow-up time. It has been documented that the effects of age, histological grade, and ER status on the survival of patients with breast cancer change over time[13,14], so it may not be comprehensive to fit only the static effects of prognostic factors.

Given covariates with time-varying effects in the field of competing risks, we proposed a dynamic effect RMTL regression model. Monte Carlo simulation was used to assess the accuracy of the coefficient estimates of the model. At the same time, we applied this model to elderly patients with early-stage breast cancer in the Surveillance, Epidemiology, and End Results (SEER) database to explore the dynamic cumulative effects and real-time effects of prognostic factors, as well as to establish a prediction model to predict mean life lost due to death from breast cancer over a period of time among patients. We hope to guide doctors to better determine the prognosis of patients, select better therapeutic regimens, and improve the survival time of patients.

## 2 Methods

### 2.1 Model Construction

Let $T$ be the time to event and $C$ be the censoring time so that the observed time is $U = \min(T, C)$. An event indicator $\varepsilon$ equals 1, 2, or 0 when the observed outcome is an event of interest, a competing event, or censoring, respectively. At the same time, let $\mathbf{Z}^* = (1, \mathbf{Z})$ denote the $n \times (p+1)$ matrix of covariates allowing an intercept term. Thus, for patient $i(i = 1, ..., n)$, the observed data include $\{U_i, \varepsilon_i, \mathbf{Z}_i^*\}$.

Let a continuous variable $l\left(0 \leq l \leq \tau, \tau \leq t_{\max}\right)$ be the pre-specified end time of follow-up, where $\tau$ is the pre-specified maximum follow-up time, and $t_{\max}$ is the natural maximum follow-up time of data. $J$ time points $l_j$ are selected from 0 to $\tau$ in ascending order and recorded as $(l_1, l_2, ..., l_J)$.

For the need of the method, we advance the end time of follow-up from $t_{\max}$ to $l_j$. Correspondingly, each patient's survival outcome will change at different pre-specified end times of follow-up. When a patient experienced the event of interest or the competing event before $l_j$, $\varepsilon(l_j)$ is equal to 1 or 2; in other cases, $\varepsilon(l_j)$ is equal to 0. $\varepsilon(s_j)$ denotes the survival outcome after restraint. Let $T(l_j) = \min(T, l_j)$ and $U(l_j) = \min(U, l_j)$ be the event time and observed time after constraint, respectively. For patient $i$, the observed data consist of $\{U_i(l_j), \varepsilon_i(l_j), \mathbf{Z}_i^*\}$.

A regression model is developed to assess the dynamic effects of covariates in RMTL:

$$g\{\mu_1(l|\ \mathbf{Z}_i^*)\} = g\{E[(l - T_i(l)) \times I(\varepsilon_i(l) = 1)]\} = \mathbf{Z}_i^* \boldsymbol{\beta}(l)$$

with link function $g(\cdot)$. $\mu_1(l|\ \mathbf{Z}_i^*)$ is the life expectancy lost due to the event of interest of patient $i$ during the $l$-year follow-up. Regression coefficients $\boldsymbol{\beta}(l) = (\boldsymbol{\beta}_1(l), \boldsymbol{\beta}_2(l), ..., \boldsymbol{\beta}_{p+1}(l))$, where $\boldsymbol{\beta}_k(l)$ is defined as $(\beta_{k0}, \beta_{k1}, \beta_{k2}) \times (1, l, l^2)$.

Because $Z_{ik}^{*} \times \boldsymbol{\beta}_k(l) = (\beta_{k0}, \beta_{k1}, \beta_{k2}) \times (Z_{ik}^{*}, l \times Z_{ik}^{*}, l^2 \times Z_{ik}^{*}) = \boldsymbol{\beta}_k \times Z_{ik}^{*}(l)$, the model can be rewritten as

$$g\{\mu_1(l \mid \boldsymbol{Z}_i^*)\} = \boldsymbol{Z}_i^*(l)\boldsymbol{\beta}.$$

Regression coefficients are estimated by solving the estimating equation

$$\Phi(\boldsymbol{\beta}) = \sum_{j=1}^{J}\sum_{i=1}^{n} I(\varepsilon_i(l_j) \neq 0) \boldsymbol{Z}_i^*(l_j) \{(l_j - T_i(l_j)) \times I(\varepsilon_i(l_j) = 1) - g^{-1}(\boldsymbol{Z}_i^*(l_j)\boldsymbol{\beta})\} = 0.$$

The model assumes the life lost of individuals with the event of interest is $(l_j - T_i(l_j)) \times I(\varepsilon_i(l_j) = 1) = l_j - T_i(l_j)$; the life lost of individuals with the competing event is $(l_j - T_i(l_j)) \times I(\varepsilon_i(l_j) = 1) = 0$; and censored observations have $I(\varepsilon_i(l_j) \neq 0) = 0$, which means they do not contribute to the estimating equation.

$E(\Phi(\boldsymbol{\beta})) \neq 0$ in the presence of censoring. However, when applying IPCW to the estimating equation, its expectation is $0^{15}$. Therefore, the estimating equation changes to

$$\Phi(\boldsymbol{\beta}) = \sum_{j=1}^{J}\sum_{i=1}^{n} \frac{I(\varepsilon_i(l_j) \neq 0)}{\hat{G}(T_i(l_j), l_j)} \boldsymbol{Z}_i^*(l_j) \{(l_j - T_i(l_j)) \times I(\varepsilon_i(l_j) = 1) - g^{-1}(\boldsymbol{Z}_i^*(l_j)\boldsymbol{\beta})\} = 0.$$

The fitted data are actually obtained by stacking $J$ datasets. The $j$-th dataset is the risk set with $l_j$-year follow-up (as shown in Figure 2). $\hat{G}(t, l_j)$ is the Kaplan-Meier estimator of the non-censoring distribution in the $j$-th dataset.

We treat the inverse probability censoring weight as a fixed value rather than a random variable[16]. Thus, the variance in the regression coefficients will not consider the variation brought by the weight. Therefore, we have $\sqrt{n}(\hat{\boldsymbol{\beta}} - \boldsymbol{\beta}) \sim N(0, \boldsymbol{A}^{-1}\boldsymbol{B}\boldsymbol{A})$, $\hat{Var}(\hat{\boldsymbol{\beta}}) = \hat{\boldsymbol{A}}^{-1}\hat{\boldsymbol{B}}\hat{\boldsymbol{A}}^{-1}$, where

$$\hat{\boldsymbol{A}} = E\left[\sum_{j=1}^{J} \boldsymbol{Z}_i^*(l_j)^{\otimes 2} h(\boldsymbol{Z}_i^*(l_j)\hat{\boldsymbol{\beta}})\right],$$

$$\hat{\boldsymbol{B}} = E\left[\sum_{j=1}^{J} \varepsilon_{ij}(\hat{\boldsymbol{\beta}})^{\otimes 2}\right],$$

$$\varepsilon_{ij}(\hat{\boldsymbol{\beta}}) = \frac{I(\varepsilon_i(l_j) \neq 0)}{\hat{G}(T_i(l_j), l_j)} \boldsymbol{Z}_i^*(l_j) \{(l_j - T_i(l_j)) \times I(\varepsilon_i(l_j) = 1) - g^{-1}(\boldsymbol{Z}_i^*(l_j)\hat{\boldsymbol{\beta}})\},$$

where $a^{\otimes 2} = aa^T$ and $h(x) = \partial g^{-1}(x)/\partial x$.

Then, we estimate regression coefficients by a generalized estimating equation, thereby correcting for data correlation.

## 2.2 Simulation

Next, we assessed the performance of the estimation of dynamic-effect RMTL regression by a simulation. We used the mean bias, mean relative bias, root mean squared error, relative standard error, and empirical coverage rate as evaluation indicators.

### 2.2.1    Data generation

First, we generated two independent variables $Z = (Z_1, Z_2)$, which were generated by an independent Bernoulli distribution. We let the subdistribution hazard function for the event of interest follow a Gompertz distribution, $\lambda_1(t|Z) = \gamma_z \exp(\rho_z t)$, where $\gamma_z$ and $\rho_z$ were set according to four strata of $(Z_1, Z_2)$.

We defined the CIF for the event of interest as $F_1(t|Z) = P(T \le t, \varepsilon = 1|Z) = 1 - \exp\{-\int_0^t \lambda_1(s|Z)ds\}$ and the CIF for the competing event as $F_2(t|Z) = P(T \le t, \varepsilon = 2|Z) = \exp(\gamma_z/\rho_z)\{1 - \exp(-t)\}$. Survival outcome was generated by Bernoulli distribution, $P(\varepsilon = 1|Z) = 1 - \exp(\gamma_z/\rho_z)$. Thus, the conditional CIF for the event of interest and the competing event were $P(T \le t|\varepsilon = 1, Z) = [1 - \exp\{-\int_0^t \lambda_1(s|Z)ds\}]/[1 - \exp(\gamma_z/\rho_z)]$ and $P(T \le t|\varepsilon = 2, Z) = 1 - \exp(-t)$, respectively. Next, we used the inverse method to generate event time $T$. Finally, we generated right censoring and determined the final observed survival outcome.

We calculated the sHRs of independent variables based on the subdistribution hazard function: $sHR(Z_1, t) = \gamma_{(1,0)} \exp(\rho_{(1,0)}t) / \gamma_{(0,0)} \exp(\rho_{(0,0)}t)$, $sHR(Z_2, t) = \gamma_{(0,1)} \exp(\rho_{(0,1)}t) / \gamma_{(0,0)} \exp(\rho_{(0,0)}t)$. We let $\rho_{(1,0)} \ne \rho_{(0,0)}$, $\rho_{(0,1)} \ne \rho_{(0,0)}$; then, the sHRs changed over time, and the proportional subdistribution hazards assumption of two variables was not satisfied.

### 2.2.2 Parameters, scenarios, and true values of simulation

Consider $(\gamma_{(0,0)}, \gamma_{(0,1)}, \gamma_{(1,0)}, \gamma_{(1,1)}) = (2.88, 1.95, 2.29, 1.55)$, $(\rho_{(0,0)}, \rho_{(0,1)}, \rho_{(1,0)}, \rho_{(1,1)}) = (-1.7, -1.4, -2.9, -2.8)$ [12]. The range of $l$ was between the 10th percentile and 95th percentile of the time of patients with the event of interest in the simulated data. Figure 3 shows that the sHRs of the independent variables changed over time. $Z_1$ was a protective factor in the early period and a risk factor in the late period, while the protective effect of $Z_2$ increased over time.

Twelve scenarios were simulated considering varying sample sizes (250, 500, 1000), proportions of exposure (both $P(Z_1 = 1)$ and $P(Z_2 = 1)$ equal to 0.25 or 0.5), and censoring rates (0.1 or 0.25). Each scenario was simulated 2000 times.

By integrating the CIF, we obtained the true value of the RMTL of each group at different $Z$, $\mu_1(l|Z) = \int_0^l F_1(t|Z)dt$. Moreover, the true values of regression coefficients were obtained by the difference in RMTL between groups: the baseline is $\beta_0(l)(= \mu_1(l|z_1=0, z_2=0))$; the cumulative effect of $Z_1$ is $\beta_1(l)(= \mu_1(l|z_1=1, z_2=0) - \mu_1(l|z_1=0, z_2=0))$; and the cumulative effect of $Z_2$ is $\beta_2(l)(= \mu_1(l|z_1=0, z_2=1) - \mu_1(l|z_1=0, z_2=0))$. Taking $l = (0.75, 1, 1.5)$, then $(\beta_0(0.75), \beta_0(1), \beta_0(1.5)) = (0.368, 0.550, 0.937)$, $(\beta_1(0.75), \beta_1(1), \beta_1(1.5)) = (-0.093, -0.15, -0.284)$, $(\beta_2(0.75), \beta_2(1), \beta_2(1.5)) = (-0.074, -0.100, -0.146)$. We found that the regression coefficient $\beta_k(l)$ was a cumulative quantity, and its absolute value increased as $l$ increased.

## 3 Results

### 3.1 Simulation Results

Table 1 and Table 2 demonstrate the accuracy of $\hat{\beta}(l)$ at different $l$. In all cases, the mean relative bias was small, in which the mean relative bias of $\beta_0(l)$ was less than 2%; the relative standard error was approximately 1; and the coverage rate was approximately 95%. Because the absolute value of true value of the regression coefficient increased with increasing $l$, it was reasonable that the mean bias increased with increasing $l$. The simulation showed that the estimation of dynamic effect RMTL regression was accurate.

With the increase in sample size, the mean bias, mean relative bias, and root mean squared error were more likely to decrease. Moreover, different censoring rates had little effect on the mean relative bias and root mean squared error.

## 3.2 Model application

In this study, we extracted data from the SEER database for elderly patients with early-stage breast cancer.

Covariates included race, age, marriage, T stage, N stage, histological grade, estrogen receptor (ER) status, progesterone receptor (PR) status, breast surgery, axillary surgery, chemotherapy, and radiotherapy[17]. Breast surgery includes mastectomy or breast-conserving surgery (BCS); while axillary surgery includes axillary lymph node dissection (ALND) and lymph node biopsy (SLNB).

We used 3892 patients diagnosed from 2000 to 2012 as a training set and another 1561 patients diagnosed from 2013 to 2015 as an externally validated set. Details of data collection and variables can be found in the supplementary material.

There were 769 deaths from breast cancer and 998 deaths from non-breast cancer in the training set, giving an approximately 55% censoring rate. The follow-up time ranged from 0.17 to 18.92 years, with a median of 8 years.

Table 3 shows the result of the static effect RMTL regression ($\tau = 10.5$ years)[12]. The regression coefficient $\beta$ indicates a cumulative difference in mean life lost during the 10.5-year follow-up due to death from breast cancer between groups of the prognostic factor. For example, $\beta_{ER} = -0.638$ showed that patients in the ER-positive group died of breast cancer 0.638 years later than those in the ER-negative group during the 10.5 years, so ER positivity was a protective factor. In general, a prognostic factor was protective when $\beta$ was negative and deleterious when $\beta$ was positive. Table 3 shows that patients with ER positivity, PR positivity, breast-conserving surgery (relative to mastectomy), and chemotherapy had a better prognosis, while patients with older age, higher T stage, higher N stage, and higher histological grade had a worse prognosis. Race, marriage, axillary surgery, and radiation therapy had no statistical significance on survival.

The significant covariates did not meet the proportional subdistribution hazards assumption, indicating that dynamic effects might exist. However, static effect RMTL regression only gives the static cumulative effect, and the real-time effect in the cumulative process cannot be known. Therefore, we fitted the proposed dynamic effect RMTL regression and used a backward stepwise approach to screen covariates. As a result, race, marriage, axillary surgery, and radiation were screened out.

Table 4 shows the results of the dynamic effect RMTL regression. Because the cumulative effect of the $k$-th prognostic factor was assumed by $\boldsymbol{\beta}_k(l) = (\beta_{k0}, \beta_{k1}, \beta_{k2}) \times (1, l, l^2)$, which would be screened by the stepwise method, the regression coefficients would include at least one of $\beta_{k0}, \beta_{k1}, \beta_{k2}$. For example, $\beta_{ER}(l) = 0.291 - 0.141l + 0.005l^2$, which was dynamic and varied with $l$. In the case of, $\beta_{ER}(4.5) \approx -0.25 \; (= 0.291 - 0.141 \times 4.5 + 0.005 \times 4.5^2)$, which means ER-negative patients lost an additional 0.25 years of life on average during the 4.5 years follow-up. Figure 4A-J shows the regression coefficients of different prognostic factors in different $l$. For example, Figure 4G shows $\beta_{ER}(l)$ (solid black line), with breast cancer deaths occurring an average of 0.25 years later (y-axis $\beta_{ER}(4.5) \approx -0.25$) in ER-positive patients than in ER-negative patients during the 4.5-year follow-up (x-axis $l = 4.5$); breast cancer deaths occurring an average of 0.42 years later (y-axis $\beta_{ER}(6.5) \approx -0.42$) in ER-positive patients than in ER-negative patients during the 6.5 years (x-axis $l = 6.5$).

In the dynamic effect RMTL regression, in addition to obtaining dynamic cumulative effects $\beta(l)$, real-time effects of prognostic factors can also be obtained by the speed of accumulation of $\beta(l)$ at different moments. We obtained real-time effects by the absolute value of the slope of the curve of $\beta(l)$, which represents the speed of accumulation. Figure 4G* and Figure 4H* show the regression coefficients (black solid lines) and the speed of accumulation (blue dashed lines) of ER and PR, respectively. In Figure 4G*, the speed of decline of $\beta(l)$ was decreasing, and the speed of decline was 0.097 when $l = 4.5$, which is in units of the difference in life lost between the positive group and negative group in the 1-year follow-up; the speed of decline was 0.077 when $l = 6.5$. Therefore, the real-time effect of ER decreased with time. In Figure 4H*, the speed of decline of $\beta(l)$ was constant, and the speeds of decline were both 0.027 when $l = 4.5$ and $l = 6.5$. Therefore, the real-time effect of PR remained unchanged with time.

We added an auxiliary line (red dotted line), which is the line between the two endpoints of the regression coefficient curve $\beta(l)$ in Figure 4A-J, to determine whether the real-time effect of the prognostic factor changed. In general, 1) when $\beta(l)$ coincided with the auxiliary line (Figure 4 A, D, F, H), the real-time effect was unchanged; 2) when $\beta(l)$ decreased with increasing $l$ (Figure 4 G, I, J), if the regression coefficient curve was below the auxiliary line, that is, the real-time effect decreased (Figure 4G, J), and conversely, the real-time effect increased (Figure 4I); and 3) when $\beta(l)$ increased with increasing $l$ (Figure 4 B, C, E), the regression coefficient curve above the auxiliary line corresponded to a decrease in the real-time effect, and conversely, it corresponded to an increase in the real-time effect (Figure 4B, C, E). Therefore, it was concluded that the real-time effects of age, stage N3 (relative to stage N1), histological grade III&IV (relative to grade II), and PR positivity were unchanged; the real-time effects of ER positivity and chemotherapy decreased; and the real-time effects of T2 (relative to T1), N2 (relative to N1), histological grade II (relative to grade I), and breast-conserving surgery increased.

In addition to exploring the dynamic cumulative effects and real-time effects of prognostic factors, another role of dynamic effect RMTL regression is providing personalized prediction for patients. Three patients were selected (see Table 5 for details). Figure 5 shows the predicted RMTL during the $l$-year follow-up of each patient, and Table 5 also shows the predicted RMTL during the 5-year and 10-year follow-up. In the case of patient A, the predicted mean life lost due to death from breast cancer was 1.5 years in the 5-year follow-up; in the decade of follow-up, the predicted value was 4.2 years. Patients B and A differed only in the choice of treatment. Compared with patient A, patient B received breast-conserving surgery and chemotherapy, and his predicted RMTL was less than that of patient A; that is, breast-conserving surgery and chemotherapy could prolong the survival time of elderly patients with early-stage breast cancer. Patient C differed from patient B in N stage and histological grade, and because patient C had lower N stage and histological grade, his predicted RMTL was lower than that of patient B.

In addition, the accuracy of prediction was evaluated by an external validation set. Figure 6 shows the C-index and prediction error when the pre-specified end time of follow-up was different[18]. The mean C-index was 0.78, indicating good discrimination of the model, and the mean prediction error was 0.47 years.

The prediction formula can be seen in Table 4, and the prediction model has been converted into a web-based prediction tool available on the web at https://m92imi-oscar-0.shinyapps.io/newapp/.

## 4 Discussion

When the effect of a prognostic factor on competing events is large, we should use a competing risks approach; otherwise, the estimate of the effect of this factor on the event of interest will be biased greatly[19]. In our data, the sHRs of age and chemotherapy on death from non-breast cancer (the competing event) were 2.486 (95% CI: 2.181 to 2.834) and 0.627 (95% CI: 0.545 to 0.722), respectively. Moreover, the number of those who experienced the competing event accounted for 26% of the total sample size and 56% of the total number of events, so it is necessary to consider competing risks in these data.

In the static effect RMTL regression, it only gives the cumulative effect during the $\tau$-year follow-up, and it is impossible to know the real-time effect in the cumulative process. In particular, this result is incomplete for covariates with time-varying effects. Additionally, for patients who have been followed up for some time, the cumulative effect from 0 to $\tau$ years is no longer applicable. In contrast, the dynamic effect RMTL regression can not only obtain the dynamic cumulative effect in the $l$-year follow-up but also explore the real-time effect. The real-time effect can help doctors and patients to have a better understanding of the prognosis of breast cancer. For example, the real-time effect of ER positivity decreased, which means its protective effect is larger in the first period and smaller in the later period, suggesting that estrogen therapy should be used as early as possible; the real-time effect of breast-conserving surgery increased, which means its protective effect is larger in the later period, suggesting that the effect of breast-conserving surgery is delayed.

Regarding the prognostic analysis of death from breast cancer, Yao used Cox regression and cause-specific Cox regression to analyze the difference in the effects of prognostic factors on breast cancer in men and women[20], and Xu used Fine-Gray regression to develop a prediction model for patients with inflammatory breast cancer[21]. However, none of these studies considered the potential time-varying effects of prognostic factors. Moreover, some studies analyzed the time-varying effects of prognostic factors[13,14,17,22], but these were the results of single-endpoint survival analysis and did not consider the impact of competing events, which may result in competing bias.

In this paper, both competing risks and time-varying effects were considered for the first time, and the real-time effects of the following prognostic factors were found to be different from the previous single-endpoint analysis results. First, in a single-endpoint analysis of breast cancer, the risk effect of stage N2 relative to stage N1 decreased over time[17]. In contrast, we found that stage N2 was also a risk factor, but the real-time effect increased over time (Figure 4C). Second, previous single-endpoint studies have shown that the deleterious effect of histological grade II relative to grade I decreased over time[13,22]. However, we found that the deleterious effect of histological grade II increased over time (Figure 4E). Third, in previous single-endpoint studies, ER positivity was a protective factor in the early period and a deleterious factor in the late period[13,22,23]. This was different from our results, which showed that the protective effect of ER positivity decreased over time (Figure 4G). Fourth, in terms of treatment, we found that patients with breast-conserving surgery had a better prognosis than those with mastectomy (Figure 4I). This is consistent with Kim's study and a meta-analysis, which showed that patients who underwent breast-conserving surgery had a higher overall survival rate than those who underwent mastectomy[24,25]. However, we further discovered that the protective effect of breast-conserving surgery increased over time (Figure 4I). Finally, chemotherapy was the protective factor, and its real-time effect decreased (Figure 4J). This is similar to Rakovich's study, which found that chemotherapy after breast-conserving surgery in patients with ductal carcinoma in situ reduced the risk of early local recurrence but not the risk of late recurrence[26].

Finally, the final dynamic RMTL model was constructed with the full dataset (see Web Table 1 in Supplementary material), and the result was similar to that constructed with the training set (Table 4).

Time-varying covariate and covariate with time-varying effect are two different types of data, which requires different statistical methods to analyze[27]. Time-varying covariate means the value of a covariate changes over time, which needs methods related to longitudinal data to analyze. While covariate with time-varying effect means the effect on the outcome is time-varying[28]. Meanwhile, covariates do not meet the proportional subdistribution hazards assumption, tending to have time-varying effect in the competing risks. Because time-varying effect is difficult to identify, we often ignore it. And then biased estimates will be obtained, and the significant effect occurring only in part of the follow-up period will be missed[29]. Among the two types of covariates, this paper focuses on the latter and proposes an extended RMTL regression model to depict time-varying effects, which also can be used in single-endpoint survival data. The extension for time-varying covariates will be the focus of our future research.

There are still some shortcomings in this study. First, the model uses IPCW. It should be noted that there are very few patients remaining at-risk at the end of follow-up, which may lead to large and unstable weights. 2) The life lost is the time lost due to death from breast cancer over a period of time (the $l$-year follow-up) rather than the reduction in total life in the traditional sense.

## 5  Conclusion

To explore the potential time-varying effects of prognostic factors under competing risks survival data, we develop a dynamic effect RMTL regression to model the stacked dataset by generalized estimating equation and IPCW technique. The simulation of regression coefficients and external validation of prediction demonstrate that dynamic effect RMTL regression is accurate in both prognosis and prediction when competing risks exist. The new model can explore dynamic cumulative effects and real-time effects of prognostic factors on a time scale, which gives clinical researchers a more comprehensive understanding of the progression of breast cancer. Moreover, time-scale-based individual prediction also allows physicians and patients to more intuitively determine the disease and choose the best treatment.

# 6 Abbreviations

**cHR:** cause-specific hazard ratio

**sHR:** subdistribution hazard ratio

**RMTL:** restricted mean time lost

**CIF:** cause-specific cumulative incidence function

**IPCW:** inverse probability of censoring weighting

**SEER:** Surveillance, Epidemiology, and End Results

**PR:** progesterone receptor status

**ER:** breast-conserving surgery

**ALND:** axillary lymph node dissection

**SLNB:** lymph node biopsy

**95%CI:** 95% confidence interval

# 8   Statements & Declarations

**Availability of data and materials:** The SEER data were available upon request to the SEER website (www.seer.cancer.gov).

**Funding:** This work was supported by the National Natural Science Foundation of China (grant numbers 82173622, 81903411) and the Guangdong Basic and Applied Basic Research Foundation (grant number 2022A1515011525).

**Acknowledgments:** The authors thank National Cancer Institute (NCI) for providing the surveillance, epidemiology, and end results (SEER) database. Any perspectives, results, or conclusions found in this paper are those of the authors.

**Competing interests:** The authors declare no conflicts of interests.

**Ethics approval and consent to participate:** Not applicable.

**Consent for publication:** Not applicable.

# 9 Table and Figure Legends

**Table Legends**

Table 1 Performance of dynamic-effect RMTL regression in the simulation when the proportion of exposure is 0.25

Table 2 Performance of dynamic-effect RMTL regression in the simulation when the proportion of exposure is 0.5

Table 3 Regression coefficients of static-effect RMTL regression ( $\tau = 10.5$ years)

Table 4 Regression coefficients of dynamic-effect RMTL regression

Table 5 The definition of three example patients

**Figure Legends**

Figure 1 Cumulative incidence curves for death from breast cancer in the ER-positive group and ER-negative group

Figure 2 Composition of the stacked dataset

Figure 3 Subdistribution hazard ratio (sHR) of two independent variables for the event of interest in the simulation

Figure 4 The curves of the regression coefficient changing over time

Figure 5 Predicted trajectories of RMTL for different patients

Figure 6 C-index and prediction error at different end times of follow-up

**Supplementary material**

Web Table 1 Regression coefficients of dynamic-effect RMTL regression (all data)

Table 1 Performance of dynamic-effect RMTL regression in the simulation when the proportion of exposure is 0.25

| N | Cen | $l$ | $\beta_0$ | | | | | $\beta_1$ | | | | | $\beta_2$ | | | | |
|---|---|---|---|---|---|---|---|---|---|---|---|---|---|---|---|---|---|
| | | | Bias ($\times 10^2$) | Rel bias | RMSE | Rel SE | Cov | Bias ($\times 10^2$) | Rel bias | RMSE | Rel SE | Cov | Bias ($\times 10^2$) | Rel bias | RMSE | Rel SE | Cov |
| 250 | 0.1 | 0.75 | -0.120 | -0.003 | 0.024 | 1.005 | 0.948 | -0.265 | 0.028 | 0.044 | 0.992 | 0.944 | 0.136 | -0.018 | 0.043 | 1.000 | 0.951 |
| | | 1 | -0.689 | -0.013 | 0.033 | 1.003 | 0.946 | 0.205 | -0.014 | 0.061 | 0.995 | 0.943 | 0.065 | -0.007 | 0.059 | 0.994 | 0.950 |
| | | 1.5 | 1.257 | 0.013 | 0.051 | 1.003 | 0.930 | 0.729 | -0.026 | 0.098 | 0.998 | 0.944 | -0.498 | 0.034 | 0.096 | 0.981 | 0.943 |
| | 0.25 | 0.75 | -0.126 | -0.003 | 0.025 | 1.035 | 0.953 | -0.233 | 0.025 | 0.047 | 0.989 | 0.940 | 0.170 | -0.023 | 0.046 | 1.001 | 0.950 |
| | | 1 | -0.699 | -0.013 | 0.035 | 1.041 | 0.952 | 0.323 | -0.022 | 0.066 | 0.993 | 0.941 | 0.163 | -0.016 | 0.065 | 0.993 | 0.950 |
| | | 1.5 | 1.245 | 0.013 | 0.059 | 1.092 | 0.949 | 1.160 | -0.041 | 0.119 | 0.987 | 0.936 | -0.187 | 0.013 | 0.119 | 0.973 | 0.938 |
| 500 | 0.1 | 0.75 | -0.068 | -0.002 | 0.017 | 0.979 | 0.945 | -0.270 | 0.029 | 0.032 | 0.979 | 0.938 | 0.034 | -0.005 | 0.030 | 1.018 | 0.953 |
| | | 1 | -0.642 | -0.012 | 0.024 | 0.986 | 0.934 | 0.208 | -0.014 | 0.043 | 0.980 | 0.939 | -0.059 | 0.006 | 0.041 | 1.016 | 0.955 |
| | | 1.5 | 1.244 | 0.013 | 0.037 | 1.010 | 0.931 | 0.777 | -0.027 | 0.070 | 0.984 | 0.940 | -0.628 | 0.043 | 0.066 | 1.005 | 0.948 |
| | 0.25 | 0.75 | -0.060 | -0.002 | 0.018 | 1.013 | 0.957 | -0.309 | 0.033 | 0.033 | 0.994 | 0.939 | 0.051 | -0.007 | 0.032 | 1.000 | 0.949 |
| | | 1 | -0.633 | -0.012 | 0.025 | 1.032 | 0.954 | 0.217 | -0.014 | 0.047 | 0.993 | 0.939 | -0.029 | 0.003 | 0.045 | 1.000 | 0.950 |
| | | 1.5 | 1.246 | 0.013 | 0.042 | 1.102 | 0.951 | 1.015 | -0.036 | 0.084 | 0.987 | 0.938 | -0.558 | 0.038 | 0.083 | 0.984 | 0.941 |
| 1000 | 0.1 | 0.75 | -0.068 | -0.002 | 0.012 | 1.000 | 0.947 | -0.387 | 0.042 | 0.022 | 1.023 | 0.949 | 0.126 | -0.017 | 0.022 | 0.989 | 0.944 |
| | | 1 | -0.646 | -0.012 | 0.017 | 1.001 | 0.928 | 0.055 | -0.004 | 0.029 | 1.024 | 0.952 | 0.066 | -0.007 | 0.030 | 0.989 | 0.947 |
| | | 1.5 | 1.235 | 0.013 | 0.028 | 1.010 | 0.921 | 0.558 | -0.020 | 0.048 | 1.025 | 0.949 | -0.451 | 0.031 | 0.048 | 0.987 | 0.950 |
| | 0.25 | 0.75 | -0.064 | -0.002 | 0.013 | 1.029 | 0.959 | -0.370 | 0.040 | 0.023 | 1.026 | 0.951 | 0.118 | -0.016 | 0.023 | 0.990 | 0.944 |
| | | 1 | -0.645 | -0.012 | 0.018 | 1.038 | 0.946 | 0.087 | -0.006 | 0.032 | 1.033 | 0.955 | 0.089 | -0.009 | 0.032 | 0.988 | 0.944 |
| | | 1.5 | 1.222 | 0.013 | 0.031 | 1.099 | 0.942 | 0.635 | -0.022 | 0.056 | 1.042 | 0.958 | -0.298 | 0.020 | 0.058 | 0.985 | 0.943 |

Abbreviations: N, the sample size; Cen, the censoring rate; $l$, end time of follow-up pre-specified; Rel bias, mean bias relative to true parameter; RMSE, root mean square error; Rel SE, mean estimated standard error/Monte Carlo empirical error; Cov, empirical coverage rate.

Table 2 Performance of dynamic-effect RMTL regression in the simulation when the proportion of exposure is 0.5

| N | Cen | $l$ | $\beta_0$ | | | | | $\beta_1$ | | | | | $\beta_2$ | | | | |
|---|---|---|---|---|---|---|---|---|---|---|---|---|---|---|---|---|---|
| | | | Bias ($\times 10^2$) | Rel bias | RMSE | Rel SE | Cov | Bias ($\times 10^2$) | Rel bias | RMSE | Rel SE | Cov | Bias ($\times 10^2$) | Rel bias | RMSE | Rel SE | Cov |
| 250 | 0.1 | 0.75 | -0.169 | -0.005 | 0.033 | 0.974 | 0.939 | -0.101 | 0.011 | 0.037 | 1.005 | 0.948 | 0.282 | -0.038 | 0.039 | 0.958 | 0.940 |
| | | 1 | -0.677 | -0.012 | 0.045 | 0.977 | 0.938 | 0.314 | -0.021 | 0.051 | 1.007 | 0.946 | 0.139 | -0.014 | 0.053 | 0.960 | 0.937 |
| | | 1.5 | 1.518 | 0.016 | 0.070 | 0.992 | 0.932 | 0.545 | -0.019 | 0.081 | 1.012 | 0.946 | -0.763 | 0.052 | 0.084 | 0.970 | 0.935 |
| | 0.25 | 0.75 | -0.179 | -0.005 | 0.034 | 0.997 | 0.947 | -0.105 | 0.011 | 0.039 | 0.999 | 0.945 | 0.329 | -0.044 | 0.041 | 0.962 | 0.942 |
| | | 1 | -0.668 | -0.012 | 0.047 | 1.002 | 0.949 | 0.324 | -0.022 | 0.055 | 1.004 | 0.955 | 0.191 | -0.019 | 0.057 | 0.961 | 0.935 |
| | | 1.5 | 1.615 | 0.017 | 0.080 | 1.027 | 0.942 | 0.613 | -0.022 | 0.094 | 1.009 | 0.954 | -0.738 | 0.051 | 0.098 | 0.966 | 0.930 |
| 500 | 0.1 | 0.75 | -0.131 | -0.004 | 0.023 | 0.997 | 0.953 | -0.219 | 0.024 | 0.026 | 0.998 | 0.946 | 0.178 | -0.024 | 0.026 | 1.000 | 0.947 |
| | | 1 | -0.619 | -0.011 | 0.031 | 1.000 | 0.948 | 0.146 | -0.010 | 0.036 | 1.001 | 0.946 | 0.010 | -0.001 | 0.036 | 0.998 | 0.948 |
| | | 1.5 | 1.627 | 0.017 | 0.050 | 1.010 | 0.940 | 0.284 | -0.010 | 0.057 | 1.009 | 0.952 | -0.928 | 0.064 | 0.059 | 0.993 | 0.943 |
| | 0.25 | 0.75 | -0.100 | -0.003 | 0.024 | 1.016 | 0.957 | -0.206 | 0.022 | 0.028 | 1.006 | 0.951 | 0.128 | -0.017 | 0.028 | 0.997 | 0.948 |
| | | 1 | -0.561 | -0.010 | 0.033 | 1.023 | 0.951 | 0.159 | -0.011 | 0.038 | 1.014 | 0.954 | -0.067 | 0.007 | 0.039 | 0.994 | 0.951 |
| | | 1.5 | 1.762 | 0.019 | 0.057 | 1.045 | 0.945 | 0.281 | -0.010 | 0.065 | 1.026 | 0.952 | -1.066 | 0.073 | 0.068 | 0.990 | 0.944 |
| 1000 | 0.1 | 0.75 | -0.190 | -0.005 | 0.016 | 1.017 | 0.956 | -0.167 | 0.018 | 0.018 | 1.027 | 0.949 | 0.289 | -0.039 | 0.019 | 0.980 | 0.944 |
| | | 1 | -0.706 | -0.013 | 0.022 | 1.023 | 0.943 | 0.218 | -0.015 | 0.025 | 1.031 | 0.952 | 0.160 | -0.016 | 0.026 | 0.987 | 0.952 |
| | | 1.5 | 1.474 | 0.016 | 0.036 | 1.037 | 0.928 | 0.376 | -0.013 | 0.040 | 1.033 | 0.954 | -0.697 | 0.048 | 0.041 | 1.001 | 0.946 |
| | 0.25 | 0.75 | -0.181 | -0.005 | 0.017 | 1.023 | 0.959 | -0.164 | 0.018 | 0.019 | 1.013 | 0.949 | 0.281 | -0.038 | 0.020 | 0.987 | 0.948 |
| | | 1 | -0.682 | -0.012 | 0.024 | 1.029 | 0.953 | 0.217 | -0.014 | 0.027 | 1.012 | 0.951 | 0.139 | -0.014 | 0.028 | 0.991 | 0.946 |
| | | 1.5 | 1.550 | 0.017 | 0.041 | 1.049 | 0.939 | 0.358 | -0.013 | 0.047 | 1.008 | 0.953 | -0.766 | 0.052 | 0.048 | 1.000 | 0.942 |

Abbreviations: N, the sample size; Cen, the censoring rate; $l$, end time of follow-up pre-specified; Rel bias, mean bias relative to true parameter; RMSE, root mean square error; Rel SE, mean estimated standard error/Monte Carlo empirical error; Cov, empirical coverage rate.

Table 3 Regression coefficients of static-effect RMTL regression ($\tau = 10.5$ years)

| Variable | Coefficient | SE | Z value | P value |
|---|---|---|---|---|
| Intercept | 1.766 | 0.480 | 3.680 | <0.001 |
| Race (ref: other) | | | | |
|   white | 0.027 | 0.174 | 0.158 | 0.874 |
|   black | -0.011 | 0.244 | -0.045 | 0.964 |
| Age (ref: 55-74) | | | | |
|   ≥75 | 0.361 | 0.125 | 2.900 | 0.004 |
| Marry (ref: other) | | | | |
|   married | -0.023 | 0.106 | -0.217 | 0.828 |
| T stage (ref: T1) | | | | |
|   T2 | 0.679 | 0.115 | 5.886 | <0.001 |
| N stage (ref: N1) | | | | |
|   N2 | 0.668 | 0.159 | 4.203 | <0.001 |
|   N3 | 1.540 | 0.249 | 6.185 | <0.001 |
| Grade (ref: grade I) | | | | |
|   II | 0.272 | 0.132 | 2.061 | 0.039 |
|   III & IV | 0.769 | 0.159 | 4.832 | <0.001 |
| ER status (ref: negative) | | | | |
|   positive | -0.638 | 0.208 | -3.070 | 0.002 |
| PR status (ref: negative) | | | | |
|   positive | -0.303 | 0.153 | -1.978 | 0.048 |
| Breast surgery (ref: mastectomy) | | | | |
|   BCS | -0.328 | 0.140 | -2.342 | 0.019 |
| Axillary surgery (ref: ALND) | | | | |
|   SLNB | -0.105 | 0.113 | -0.933 | 0.351 |
| Chemotherapy (ref: no) | | | | |
|   yes | -0.320 | 0.123 | -2.603 | 0.009 |
| Radiation (ref: no) | | | | |
|   yes | -0.527 | 0.400 | -1.317 | 0.188 |

Note: Except for race, marriage, breast surgery, and axillary surgery, none of the other variables met the proportional subdistribution hazards assumption.

The regression formula of the static effect RMTL model is as follows:

$$\begin{aligned}
RMTL = &\ 1.766 + 0.027 \times I(race = white) - 0.011 \times I(race = black) + 0.361 \times I(age \geq 75) \\
&- 0.023 \times I(marry = married) + 0.679 \times I(T\ stage = T2) + 0.668 \times I(N\ stage = N2) \\
&+ 1.54 \times I(N\ stage = N3) + 0.272 \times I(grade = \mathrm{II}) + 0.769 \times I(grade = \mathrm{III\ \&\ IV}) \\
&- 0.638 \times I(ER = positive) - 0.303 \times I(PR = positive) - 0.328 \times I(breast\ surgery = BCS) \\
&- 0.105 \times I(axillary\ surgery = SLNB) - 0.32 \times I(chemotherapy = yes) \\
&- 0.527 \times I(radiation = yes)
\end{aligned}$$

Table 4 Regression coefficients of dynamic-effect RMTL regression

（2.5 years ≤ $l$ ≤ 10.5 years）

| Variable | Time function | Coefficient | SE | Z value | P value |
|---|---|---|---|---|---|
| Intercept | 1 | -0.237 | 0.077 | -3.071 | 0.002 |
|  | $l$ | 0.140 | 0.026 | 5.311 | <0.001 |
| Age (ref: 55-74) | | | | | |
| age 75+ | 1 | -0.104 | 0.035 | -2.951 | 0.003 |
|  | $l$ | 0.047 | 0.012 | 4.062 | <0.001 |
| T stage (ref: T1) | | | | | |
| T2 | $l^2$ | 0.006 | 0.001 | 7.010 | <0.001 |
| N stage (ref: N1) | | | | | |
| N2 | $l^2$ | 0.006 | 0.001 | 4.865 | <0.001 |
| N3 | 1 | -0.404 | 0.082 | -4.919 | <0.001 |
|  | $l$ | 0.177 | 0.026 | 6.701 | <0.001 |
| Grade (ref: grade I) | | | | | |
| II | $l^2$ | 0.002 | 0.001 | 2.881 | 0.004 |
| III & IV | 1 | -0.233 | 0.044 | -5.317 | <0.001 |
|  | $l$ | 0.093 | 0.015 | 6.383 | <0.001 |
| ER status (ref: negative) | | | | | |
| positive | 1 | 0.291 | 0.061 | 4.731 | <0.001 |
|  | $l$ | -0.141 | 0.024 | -5.823 | <0.001 |
|  | $l^2$ | 0.005 | 0.001 | 3.970 | <0.001 |
| PR status (ref: negative) | | | | | |
| positive | $l$ | -0.027 | 0.010 | -2.694 | 0.007 |
| Breast surgery (ref: mastectomy) | | | | | |
| BCS | $l^2$ | -0.003 | 0.001 | -2.687 | 0.007 |
| Chemotherapy (ref: no) | | | | | |
| yes | 1 | 0.160 | 0.044 | 3.648 | <0.001 |
|  | $l$ | -0.092 | 0.017 | -5.329 | <0.001 |
|  | $l^2$ | 0.004 | 0.001 | 3.166 | 0.002 |

Note: The regression formula for the dynamic-effect RMTL regression is as follows:

$$\begin{aligned} RMTL(l) =& -0.237 + 0.14l + (-0.104 + 0.047l) \times I(age \geq 75) + 0.006l^2 \times I(T\ stage = T2) \\ &+ 0.006l^2 \times I(N\ stage = N2) + (-0.404 + 0.177l) \times I(N\ stage = N3) + 0.002l^2 \times I(grade = \text{II}) \\ &+ (-0.233 + 0.093l) \times I(grade = \text{III \& IV}) + (0.291 - 0.141l + 0.005l^2) \times I(ER = positive) \\ &- 0.027l \times I(PR = positive) - 0.003l^2 \times I(breast\ surgery = BCS) + (0.16 - 0.092l + 0.004l^2) \\ &\times I(chemotherapy = yes). \end{aligned}$$

Table 5 The definition of three example patients

| Patient | N stage | grade | surgery | chemotherapy | $RMTL(5)$ | $RMTL(10)$ |
|---|---|---|---|---|---|---|
| A | N3 | IV | mastectomy | no | 1.5 | 4.2 |
| B | N3 | IV | BCS | yes | 1.2 | 3.5 |
| C | N1 | I | BCS | yes | 0.5 | 1.5 |

Note: Three patients were over 75 years of age, ER negative, PR negative, and had T2 stage cancer. $RMTL(5)$ is the predicted mean life lost due to death from breast cancer during the five-years follow-up; $RMTL(10)$ is the corresponding predicted value during the decade of follow-up.

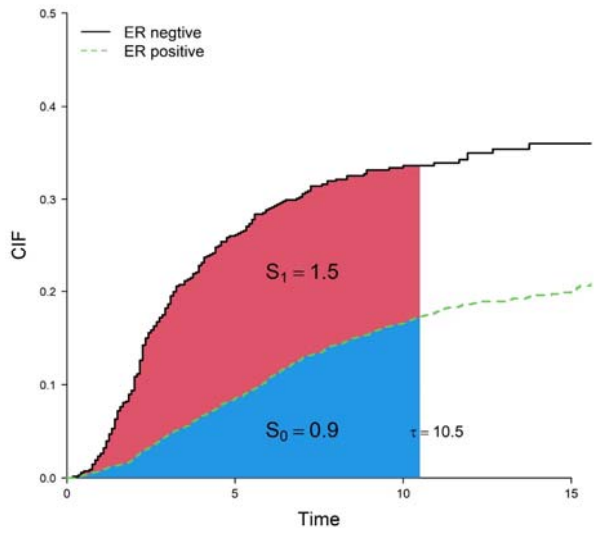

Figure 1 Cumulative incidence curves for death from breast cancer in the ER-positive group and ER-negative group

Note: $S_0, S_1$ correspond to the blue and red areas, respectively.

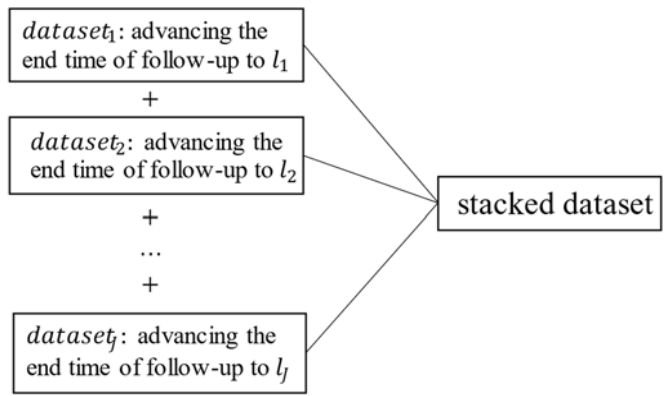

Figure 2 Composition of the stacked dataset

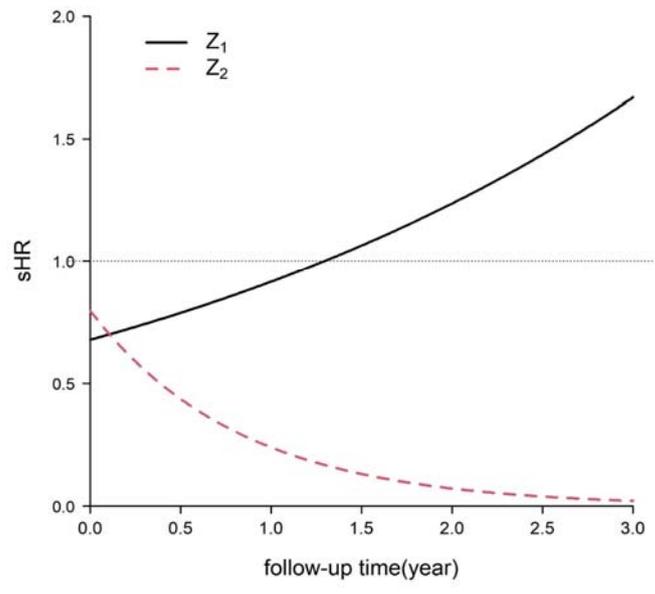

Figure 3 Subdistribution hazard ratios (sHRs) of two independent variables for the

event of interest in the simulation

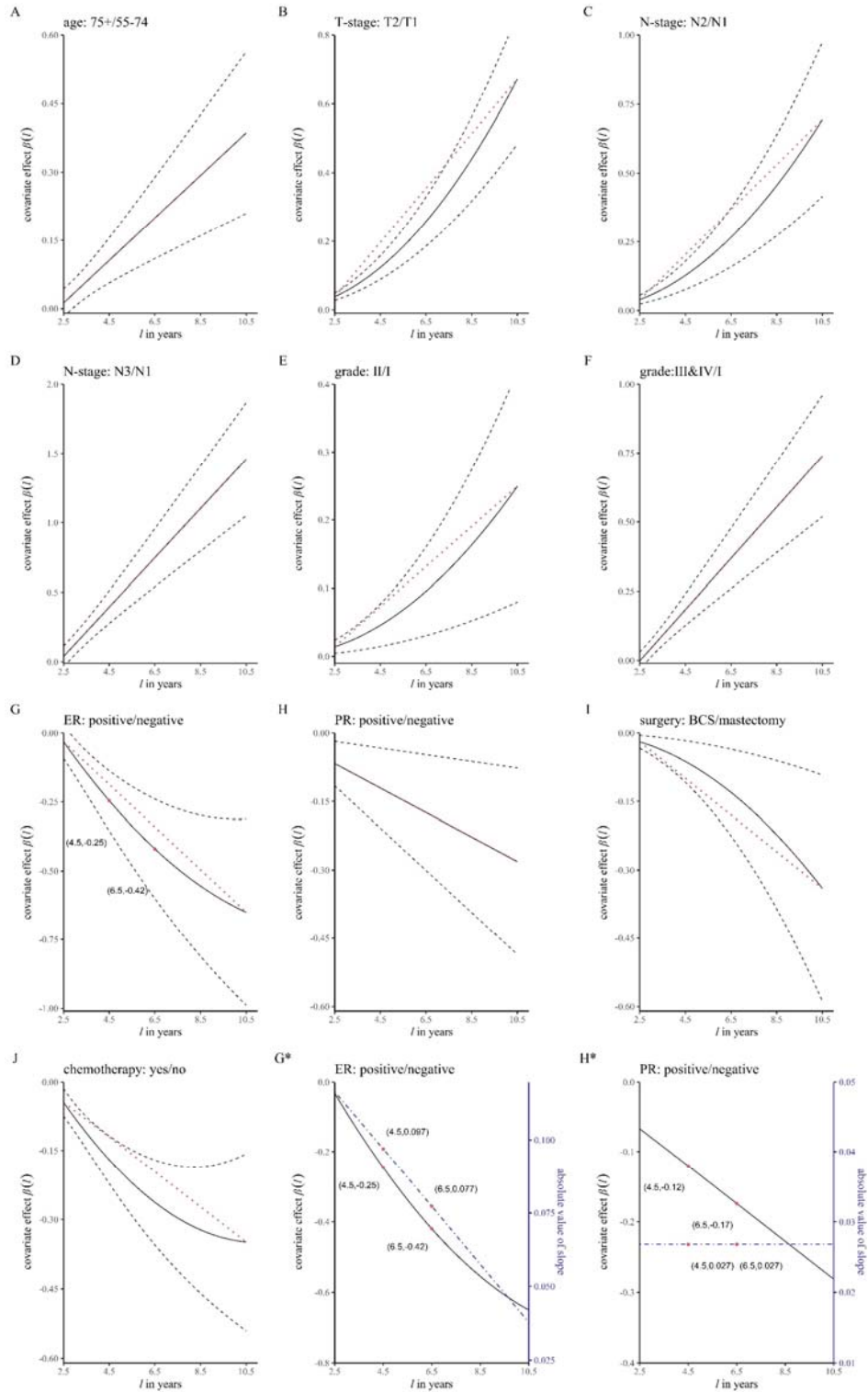

Figure 4 The curves of the regression coefficient changing over time
Note: The letters A-J represent different variables. The black solid line represents the regression coefficient $\beta(l)$, the black dashed line represents the 95% confidence interval of $\beta(l)$, and the red dotted line is an auxiliary line (straight line between two endpoints of $\beta(l)$), which is used to judge whether $\beta(l)$ is a curve. In G* and H*, the blue dashed line corresponds to the right coordinate and is the absolute value of the slope of $\beta(l)$.

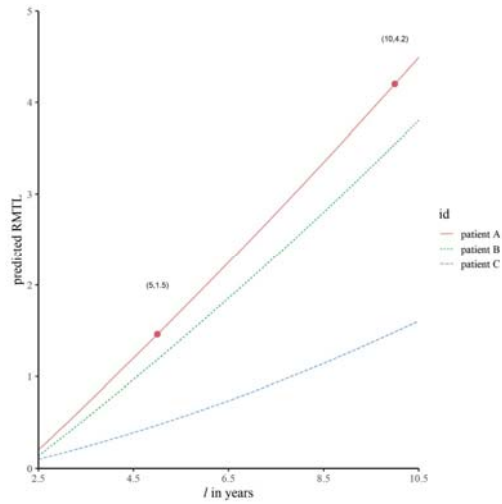

Figure 5 Predicted trajectories of RMTL for different patients

Note: The predicted mean life lost of patient A due to death from breast cancer was 1.5 years in the 5-year follow-up; in the decade of follow-up, the corresponding predicted value was 4.2 years.

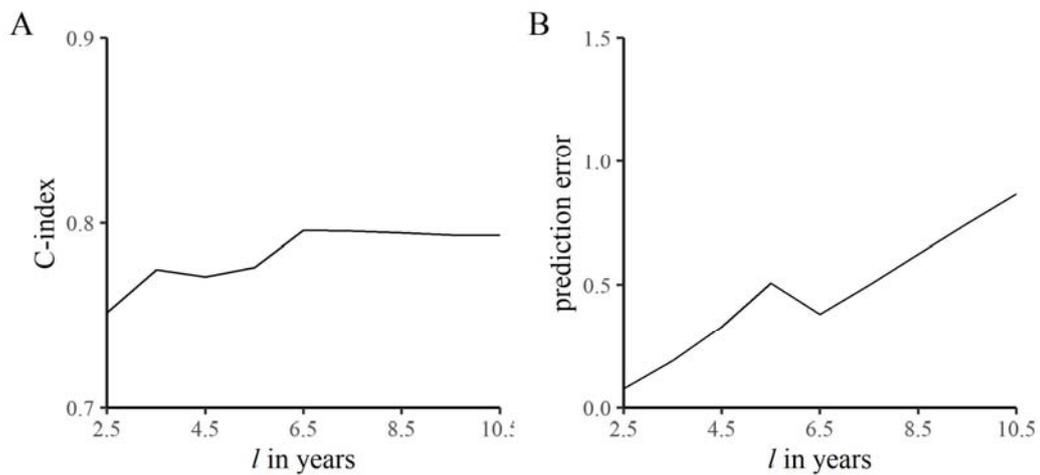

Figure 6 C-index and prediction error at different end times of follow-up

Note: The C-index refers to the accuracy of the model in predicting the sequence of occurrence of death from breast cancer in the $l$-year follow-up. Prediction error refers to the error between the predicted mean life lost due to death from breast cancer and the corresponding true value in the $l$-year follow-up.